\documentclass[prd,preprint, tightenlines,superscriptaddress,showpacs,byrevtex]{revtex4-1}

\usepackage{graphicx}
\usepackage{subfigure}
\usepackage{overpic}
\usepackage{epsfig}
\usepackage{dcolumn}
\usepackage{bm}
\usepackage{color}
\usepackage{siunitx}

\newcommand{\gev}{\rm GeV}

\newcommand{\gevcs}{{\rm GeV}/c^2}
\newcommand{\mev}{\rm MeV}
\newcommand{\mevcs}{{\rm MeV}/c^2}

\newcommand{\ev}{\rm eV}

\newcommand{\nb}{\si{\nano\barn}}
\newcommand{\pb}{\si{\pico\barn}}
\newcommand{\fb}{\si{\femto\barn}}
\newcommand{\infb}{\fb^{-1}}
\newcommand{\inpb}{\pb^{-1}}

\newcommand{\MMS}{M_{\rm miss}^2}

\newcommand{\yns}{\Upsilon(nS)}

\newcommand{\lum}{{\cal L}}
\newcommand{\eff}{\varepsilon}
\newcommand{\BR}{{\cal B}}

\newcommand{\piz}{\pi^0}
\newcommand{\etap}{\eta^{\prime}}

\newcommand{\jpsi}{J/\psi}

\newcommand{\phit}{\phi(1680)}
\newcommand{\phiy}{\phi(2170)}
\newcommand{\EE}{e^+e^-}

\newcommand{\GG}{\gamma\gamma}
\newcommand{\gl}{\gamma_l}
\newcommand{\gh}{\gamma_h}
\newcommand{\pp}{\pi^+\pi^-}
\newcommand{\ppp}{\pi^+\pi^-\pi^0}
\newcommand{\kk}{K^+K^-}

\newcommand{\gisr}{\gamma_{\rm ISR}}

\newcommand{\evis}{E_{\rm vis}}

\newcommand{\cm}{\si{\centi\metre}}

\newcommand{\reduline}{\bgroup\markoverwith
{\textcolor{red}{\rule[0.5ex]{2pt}{0.4pt}}}\ULon}

\newcommand{\beq}{\begin{equation}}
\newcommand{\eeq}{\end{equation}}
\newcommand{\beqar}{\begin{eqnarray}}
\newcommand{\eeqar}{\end{eqnarray}}
\newcommand{\bitm}{\begin{itemize}}
\newcommand{\eitm}{\end{itemize}}

\def\Journal#1#2#3#4{{#1} {\bf #2}, #3 (#4)}
\def\IJMP{Int. J. Mod. Phys. A}

\def\NIMA{Nucl. Instrum. Methods A}
\def\NPA{Nucl. Phys. A}

\def\PLB{Phys. Lett. B}
\def\PRL{Phys. Rev. Lett.}
\def\PRD{Phys. Rev. D}
\def\PRP{Phys. Rep.}

\def\EPJC{Eur. Phys. J. C}

\begin{document}


\title{
\quad\\[1.0cm]
Study of $\EE\to\eta\phi$ via Initial State Radiation at Belle}

\author{The Belle Collaboration}

\date{\today}

\begin{abstract}

Using $980~\infb$ of data collected on and around the $\Upsilon(nS)(n=1,2,3,4,5)$ resonances with the Belle detector at 
the KEKB collider, we measure the cross section of $\EE\to \eta\phi$ from threshold to $3.95~\gev$ via initial-state 
radiation. From a multi-parameter fit assuming $\phiy$ exists in the $\eta\phi$ final state according to previous 
measurement by BESIII, the resonant parameters of $\phit$ are determined to be $m_{\phit} = (1683 \pm 7 \pm 9)~\mevcs$ 
(statistical and systematic errors, respectively), $\Gamma_{\phit} = (149 \pm 12 \pm 13)~\mev$ and, depending on the 
possible presence of interfering resonances, $\Gamma^{\EE}_{\phit}\cdot \BR[\phit\to\eta\phi] = (122 \pm 6 \pm 
13)~\ev$, $(219 \pm 15 \pm 18)~\ev$, $(163 \pm 11 \pm 13)~\ev$ or $(203 \pm 12 \pm 18)~\ev$. The branching fraction of 
$\phit\to\eta\phi$ decay is determined to be approximately 20\%. Additionally, the branching fraction for 
$\jpsi\to\eta\phi$ is measured to be $(7.1\pm 1.0 \pm 0.5)\times 10^{-4}$. However, there is no significant observed 
$\phiy$ signal in the $\eta\phi$ final states in this analysis, and correspondingly the upper limit for 
$\Gamma^{\EE}_{\phiy}\cdot \BR(\phiy\to\eta\phi)$ is determined to be either $0.17~\ev$ (for two fits), or  $18.6~\ev$ 
(remaining two fits), at 90\% confidence level.

\end{abstract}


\maketitle

\section{Introduction}

Quarkonium and quarkonium-like states play an important role in understanding Quantum Chromodynamics (QCD), which is the 
generally accepted theory for strong interactions between quarks and gluons. However, there are no first-principles 
methods to derive the spectrum and properties of hadrons from the QCD Lagrangian. Alternatively, the more 
phenomenological Quark Model is used comprehensively~\cite{review}. Although hadrons with multiple quarks ($n>3$), with 
only gluons, or with bound hadrons, etc., are allowed according to QCD, only recently have accordant candidates been 
identified. Since the discovery of $X(3872)$ in 2003 by the Belle experiment~\cite{x3872}, dozens of new states have 
been observed by Belle, BaBar, BESIII, CLEOc, LHCb, etc. However, these new states do not easily fit into the hadronic 
spectrum derived from the Quark Model, indicating that new types of hadrons may have already been observed. For 
example, the charged charmonium-like states, such as $Z_c(3900)$~\cite{z3900}, $X(4020)^{\pm}$~\cite{z4020} and 
$X(4055)^{\pm}$~\cite{z4055}, are generally interpreted as exotic states.

Hadronic transitions have contributed significantly to the discoveries of quarkonium(-like) states, such as the 
$Y(4260)$ in $\EE\to \pp\jpsi$ via initial-state radiation (ISR) by the BaBar experiment~\cite{babay4260}. In searching 
for an $s\bar{s}$ version of the $Y(4260)$, the $Y(2175)$ (now called `$\phi(2170)$') was discovered in $\EE\to\pp\phi$ 
via ISR by BaBar~\cite{y2175_babar}, and later confirmed by Belle~\cite{y2175_belle}. There are several interpretations 
of the $\phi(2170)$, such as a regular $s\bar{s}$ meson~\cite{phi_2d, phi_3s}, an $s\bar{s}g$ hybrid~\cite{phi_ssg}, a 
tetraquark state~\cite{tetra_1, tetra_2, tetra_3}, a $\Lambda \bar{\Lambda}$ bound state~\cite{bound_1, bound_2, 
bound_3, bound_4}, an $S$-wave threshold effect~\cite{threshold}, or a three-meson system $\phi KK$~\cite{phikk}. In a 
recent lattice QCD calculation~\cite{lattice-qcd}, the properties of the lowest two states comply with those of 
$\phi(1020)$ and $\phit$, but with no obvious correspondence to the $\phiy$. In searching for $\phi(2170)$ in other 
hadronic transitions, BaBar studied the $\EE\to\eta\phi$ process via ISR using a $232~\infb$ data sample and found 
several hundreds of $\eta\phi$ signal events, among which hints of an excess were observed around $2.1~\gevcs$ 
~\cite{etaphi_babar}. Assuming these hints correspond to bound $\phi^{''}$ state, BaBar estimated the mass 
$M_{\phi^{''}} = (2125\pm 22\pm 10)~\mevcs$, width $\Gamma_{\phi^{''}} =  (61\pm 50\pm 13)~\mev$ and product of the 
partial width times branching fraction $\Gamma^{\EE}_{\phi^{''}} \BR(\phi^{''}\to \phi\eta) = (1.7\pm 0.7 \pm 1.3)~\ev$. 
(Hereinafter, quoted uncertainties are statistical systematic, respectively.) The CMD-3 experiment measured the process 
$\EE\to KK\eta$ from 1.59 to $2.007~\gev$ and found it is dominated by the $\eta\phi$ contribution, and then calculated 
the contribution to the anomalous magnetic moment of muon: $\alpha_\mu^{\eta\phi}(E<1.8~\gev) = (0.321\pm 0.015\pm 
0.016)\times 10^{-10}$, $\alpha_\mu^{\eta\phi} (E<2.0~\gev) = (0.440\pm 0.015\pm 0.022)\times 10^{-10}$ 
~\cite{etaphi_cmd3}. Recently, BESIII measured $\EE\to \phi\etap$ with a data sample taken at center of mass (CM) 
energies ($\sqrt{s}$) ranging from $2.05$ to $3.08~\gev$ and observed a resonance near $2.17~\gev$ with a statistical 
significance exceeding $10\sigma$~\cite{etapphi}. If both of these correspond to decays of the $\phiy$, one could infer 
the ratio $\BR(\phiy\to\phi\eta)/\BR(\phiy\to\phi\etap) = 0.23\pm 0.10 \pm 0.18 $, which is smaller than the prediction 
of $s\bar{s}g$ hybrid models by several orders of magnitude. However, due to limited statistics, the uncertainty in 
$\Gamma^{\EE}_{\phi^{''}}\BR(\phi^{''}\to \phi\eta)$ from BaBar is large. BESIII also measured the Born cross section of 
$\EE\to\eta\phi$ and determined the $\phi(2170)$ parameters to be $m_{\phiy} = (2163.5 \pm 6.2 \pm 3.0)~\mevcs$, 
$\Gamma_{\phiy} = (31.1^{+21.1}_{-11.6} \pm 1.1)~\mev$, and $\Gamma^{\EE}_{\phiy}\BR(\phiy\to \phi\eta) = 
(0.24^{+0.12}_{-0.07}) ~\ev$ or $(10.11^{+3.87}_{-3.13})~\ev$ ~\cite{etaphi_bes3}. The signal significance of $\phiy$ 
is determined to be $6.9\sigma$. In that analysis, BESIII used, as input, the cross section of $\EE\to\eta\phi$ below 
$2.0~\gev$ (dominated by the $\phit$ signal) measured by BaBar~\cite{etaphi_babar} in the determination of the $\phiy$ 
resonant parameters.

In this article, we report a study of the $\EE \to\eta\phi$ process via ISR with the Belle detector ~\cite{Belle} at the 
KEKB asymmetric-energy $\EE$ collider~\cite{KEKB}. The integrated luminosity used in this analysis is $980~\infb$, of 
which $\sim$70\% were collected at the $\Upsilon(4S)$ resonance, with the remainder accumulated either at the other 
$\yns(n=1, 2, 3, 5)$ resonances or at $\sqrt{s}$ lower than the $\Upsilon$ resonances by tens of $\mev$. This data 
sample is much larger than the one used in the previous analysis~\cite{etaphi_babar}. We scan the $\phit\to\eta\phi$ 
final state over the energy interval from $1.7~\gevcs$ to $3.95~\gevcs$, which also covers the signal regions for 
$\phiy$ and $\jpsi$. The well improved precision of the cross section of $\EE\to\eta\phi$ will be helpful to calculated 
the $\alpha_\mu^{\eta\phi}$~\cite{alpha_mu}. The $\phi$ is reconstructed from its decay to $\kk$ final state, and the 
$\eta$ is reconstructed from its decay to either the $\GG$ or $\ppp$ final states.

\section{The Belle Detector and Monte Carlo (MC) simulation}

The Belle detector is a large-solid-angle magnetic spectrometer consisting of a silicon vertex detector, a 50-layer 
central drift chamber, an array of aerogel threshold Cherenkov counters, a barrel-like arrangement of time-of-flight 
scintillation counters, and an electromagnetic calorimeter (ECL) comprised of CsI(Tl) crystals located inside a 
superconducting solenoid coil that provides a 1.5T magnetic field. An iron flux return located outside of the coil is 
instrumented to detect $K^0_{\rm L}$ mesons and to also identify muons. With the origin of the coordinate system 
defined as the nominal interaction point, the $z$ axis is aligned with the direction opposite the $e^+$ beam and is 
parallel to the direction of the magnetic field within the solenoid. The $y$ axis is vertical upward, and the $x$ axis 
is horizontal and completes the right-handed coordinate frame. The polar angle $\theta$ and azimuthal angle $\phi$ are 
measured relative to the positive $z$ and $x$ axes, respectively.

The {\sc phokhara} event generator~\cite{phokhara} is used to simulate the process $\EE \to \eta\phi$ via ISR for 
optimization of selection criteria and the efficiency estimation. One or more ISR photons ($\gisr$) are emitted before 
forming a resonance $Y$, which then decays to $\eta\phi$ with $\phi\to\kk$ and $\eta\to\ppp$ or $\GG$. In the generator, 
the resonance $Y$ could be $\phit$, $\phiy$, $\jpsi$ or a particle with mass fixed to a value between 1.6 and 
$4.0~\gevcs$ and width fixed to zero. Since the $\phit$ dominates the $\eta\phi$ final states, we use the MC sample of 
$\phit$ as the nominal signal MC sample. A GEANT3-based MC simulation~\cite{geant3} is used to simulate the Belle 
detector response. 

\section{Event selection criteria}

To study the $\eta\phi$ final states, a $\phi$ candidate is reconstructed from a $\kk$ pair and an $\eta$ candidate is 
reconstructed in either the $\GG$ or $\ppp$ ($\piz\to\GG$) modes. Hereinafter, the reconstruction channel with 
$\eta\to\GG$ is called  the ``$\GG$ mode", and the three-pion mode is referred to as the ``$\ppp$ mode." For a candidate 
event, we require two (four)  well-measured charged tracks with zero net charge for the $\GG$ ($\ppp$) mode. A 
well-measured charged track is defined as one having impact parameters with respect to the interaction point satisfying 
$dr < 1.5~\cm$ in the $r-\phi$ plane and $|dz|<5~\cm$ in the $r-z$ plane, respectively. For each charged track, 
information from different detector subsystems is combined to form a likelihood $\mathcal{L}_i$ for each putative 
particle species ($i$)~\cite{pid}. Tracks with $\mathcal{R}_K = \frac{\mathcal{L}_K} {\mathcal{L}_K + \mathcal{L}_\pi} 
> 0.6$ are identified as kaons, while those with $\mathcal{R}_K < 0.4$ are identified as pions, with an efficiency of 
about 95\% for $K-\pi$ separation.

Each photon candidate is a cluster in the ECL that is unmatched to the extrapolated trajectories of any charged tracks. 
The photon with the highest energy is identified to be $\gisr$. In the reconstruction of $\piz$ candidates, the energy 
of a photon candidate is required to have $E_\gamma > 25~\mev$ in the barrel ($\cos\theta \in  [-0.63, 0.85]$) and 
$E_\gamma > 50~\mev$ in the endcaps ($\cos\theta \in [-0.91, -0.65]\cup [0.85, 0.98]$). The $M_{\GG}$ mass resolution is 
about $6~\mevcs$, and the signal region of the $\piz$ is defined to be  $120 < M_{\GG} < 150~\mevcs$ with $\chi^2(\piz)< 
25$ (the $\chi^2$ value returned for the mass fit to each $\piz$ candidate). Events with $\gamma\to \EE$ conversions are 
removed by requiring $\mathcal{R}_e < 0.75$ for the $\pp$ tracks from $\eta$ decays. In this case, the particle 
identification variable for electron/positron in conversion products is defined as $\mathcal{R}_e\equiv 
\mathcal{L}_e/(\mathcal{L}_e+\mathcal{L}_{\rm hadrons}$). In the reconstruction of $\eta\to\GG$, two photon candidates 
are required, each with energy satisfying $E_{\gl} > 120~\mev$ and $E_{\gh} > 350~\mev$, where the subscript $l$ ($h$) 
signifies the lower (higher) energy photon in the laboratory system. The efficiency of the energy requirement is  
$(96.6 \pm 0.1)\%$ (statistical error only), as determined from signal MC simulation. 

\begin{figure}[tbp]
\includegraphics[width=0.45\textwidth]{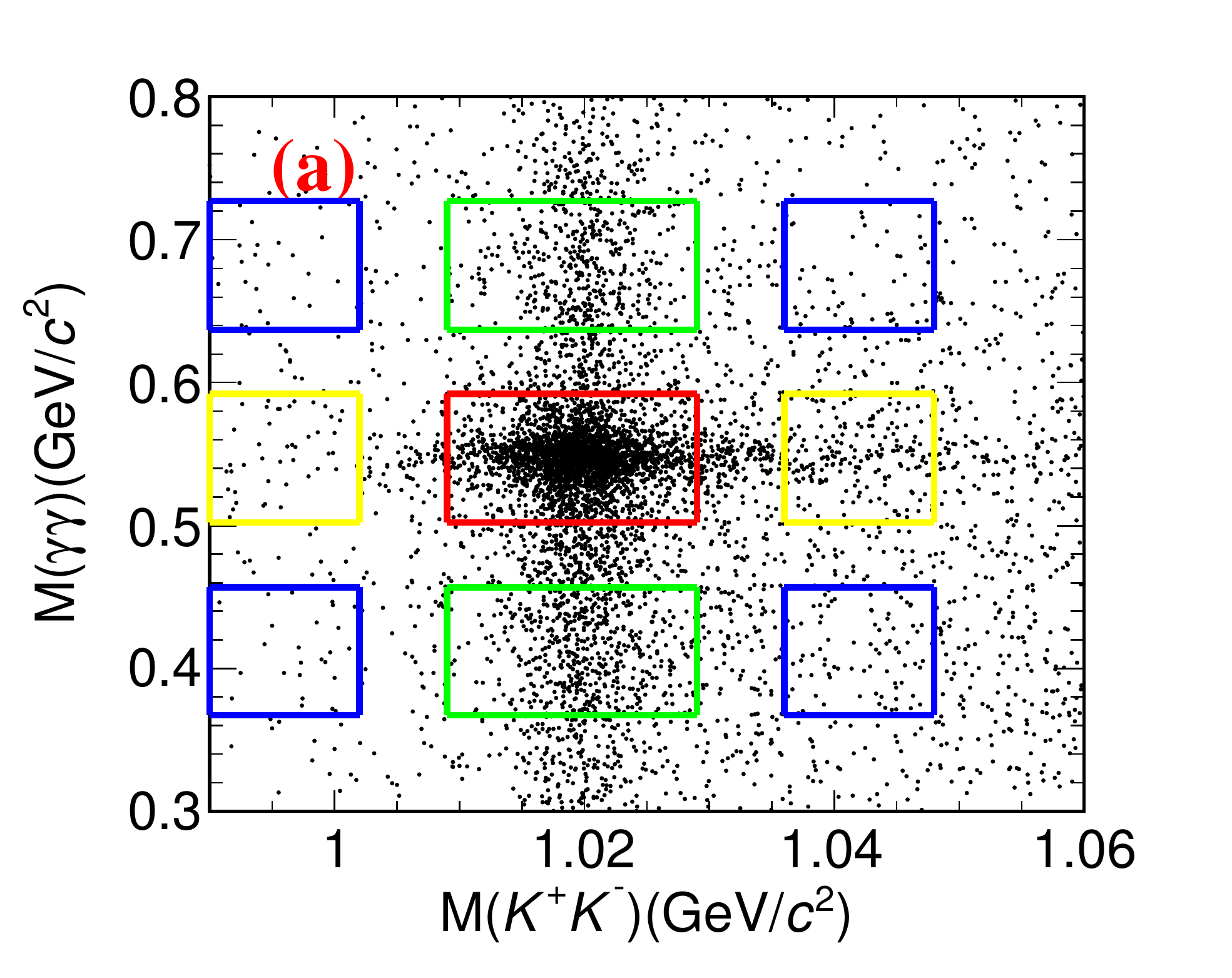}
\includegraphics[width=0.45\textwidth]{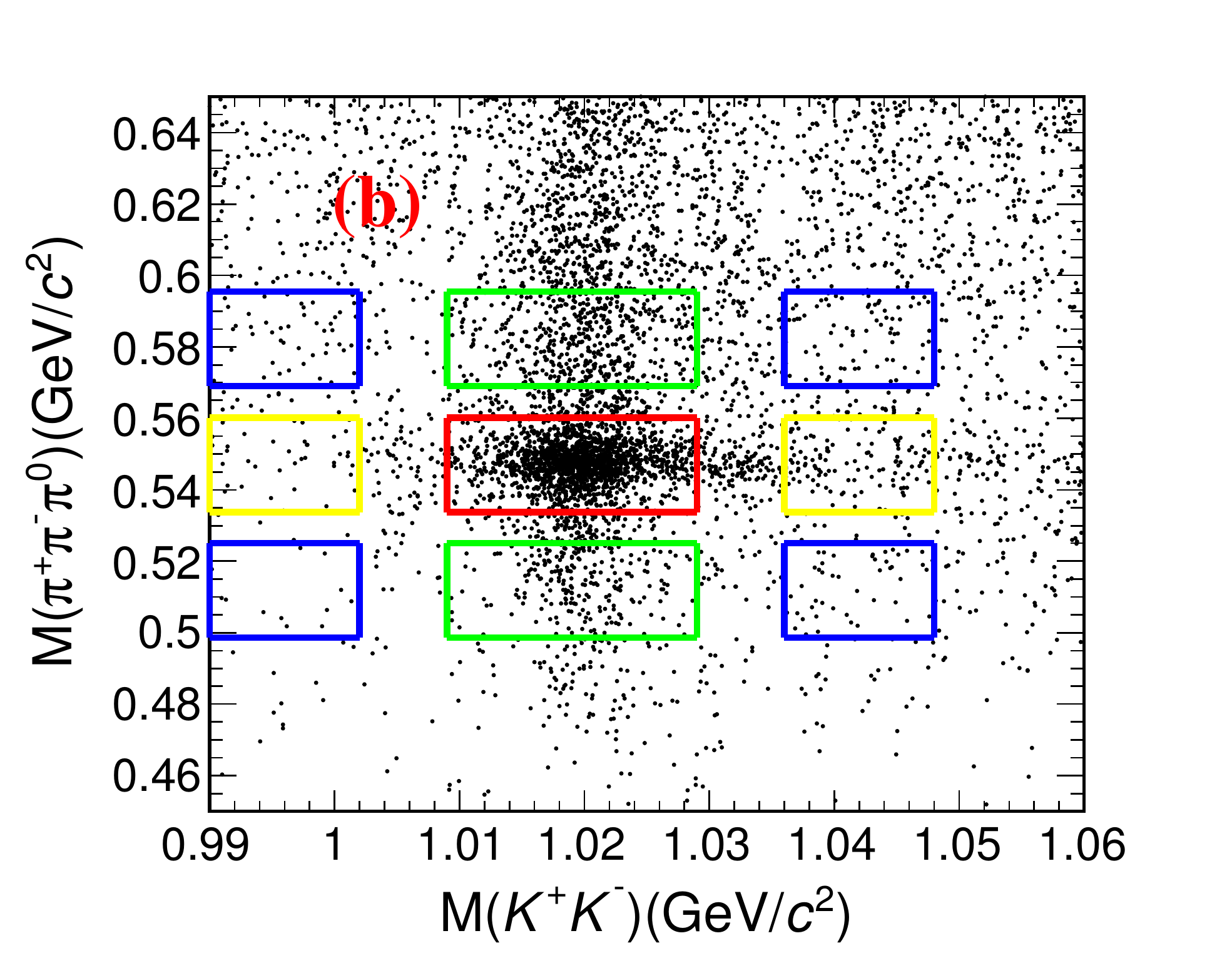}
\caption{Invariant mass distributions of (a) $\kk$ versus $\GG$ and (b) $\kk$ versus $\ppp$ for the selected $\ppp\kk$ 
or $\GG\kk$ candidates having $\eta\phi$ invariant mass below $3.5~\gevcs$. The box in the center of each plot shows 
the $\eta\phi$ signal region, while the surrounding boxes show the sideband regions, defined according to the scheme described in the text.} 
\label{mkkmeta}
\end{figure}

The scatter plots displaying the dikaon ($M_{\kk}$) invariant mass versus the $\ppp$ invariant mass ($M_{\ppp}$), or the 
$\gl\gh$ invariant mass ($M_{\GG}$) are shown in Fig.~\ref{mkkmeta}. A $\kk$ pair is treated as a $\phi$ candidate if 
$|M_{\kk}-m_{\phi}| < 12~\mevcs$ (the mass resolution is $\sim 4~\mevcs$), where $m_\phi$ is the $\phi$ nominal 
mass~\cite{PDG}. This mass interval requirement for the $\phi$ retains $(97.1\pm0.6)\%$ of $\phi$ candidates in data and 
$(97.4\pm 0.1)\%$ in the signal MC simulation, respectively. The lower and upper $\phi$ mass sidebands are defined to be 
$0.990 < M_{\kk} < 1.002~\gevcs$ and $1.036 < M_{\kk} < 1.048~\gevcs$. A fit to the $M_{\ppp}$ or $M_{\GG}$ distribution 
with a Gaussian function for the $\eta$ signal, and a smooth second-order polynomial function for background yields a 
mass resolution of $\sigma_{\ppp} = 4.2~\mevcs$ in the $\ppp$ mode and $\sigma_{\GG} = 11.3~\mevcs$ in the $\GG$ mode. 
We define the $\eta$ signal mass interval by $|M_{\ppp/\GG} - m_\eta| < 3\sigma_{\ppp/\GG}$, and the sideband regions 
are defined by $|M_{\ppp} - m_\eta \pm 9\sigma_{\ppp/\GG} | < 3\sigma_{\ppp/\GG}$, where $m_\eta$ is the nominal $\eta$ 
mass ~\cite{PDG}. The central (surrounding) rectangles of Fig.~\ref{mkkmeta} show the $\eta\phi$ signal (sideband) 
regions. With $S1$, $S2$ and $S3$ representing the sum of the events in the two adjacent horizontal ($S1$) and vertical 
($S2$) sideband boxes, and ($S3$) the four diagonal sideband boxes relative to the signal box, the normalization of the 
two-dimensional (2D) sidebands is given by $S = a\cdot S1 +b\cdot S2 - ab\cdot S3$, where $a = 0.84\pm 0.05$ and $b = 
0.52\pm 0.03$ are the appropriate areal scale factors, according to the $M_{\kk}$ and $M_{\ppp/\GG}$ distributions. 
These 2D sidebands are used to estimate the background level in the $\eta\phi$ signal region.

\begin{figure}[tbp]
\includegraphics[width=0.3\textwidth]{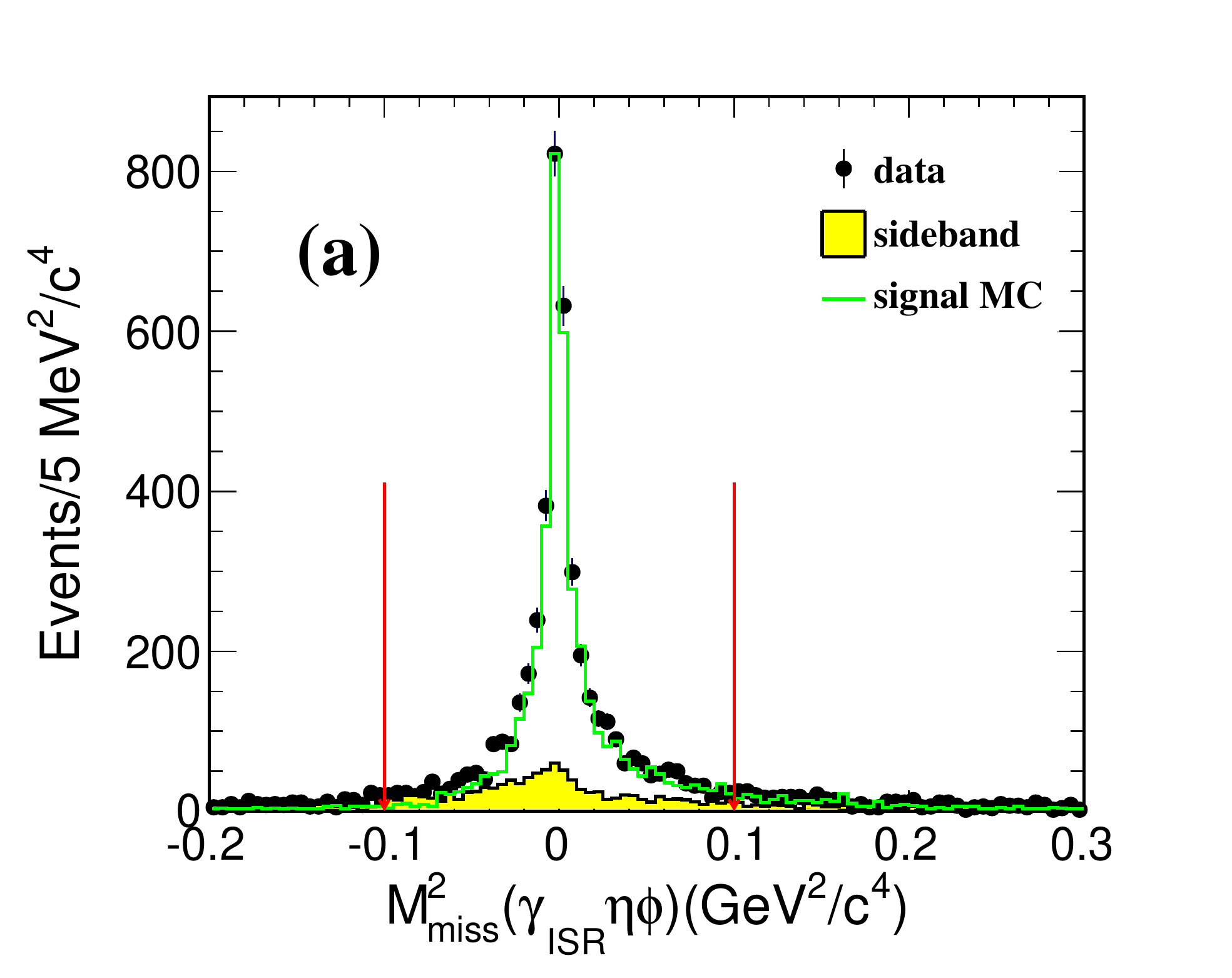}
\includegraphics[width=0.3\textwidth]{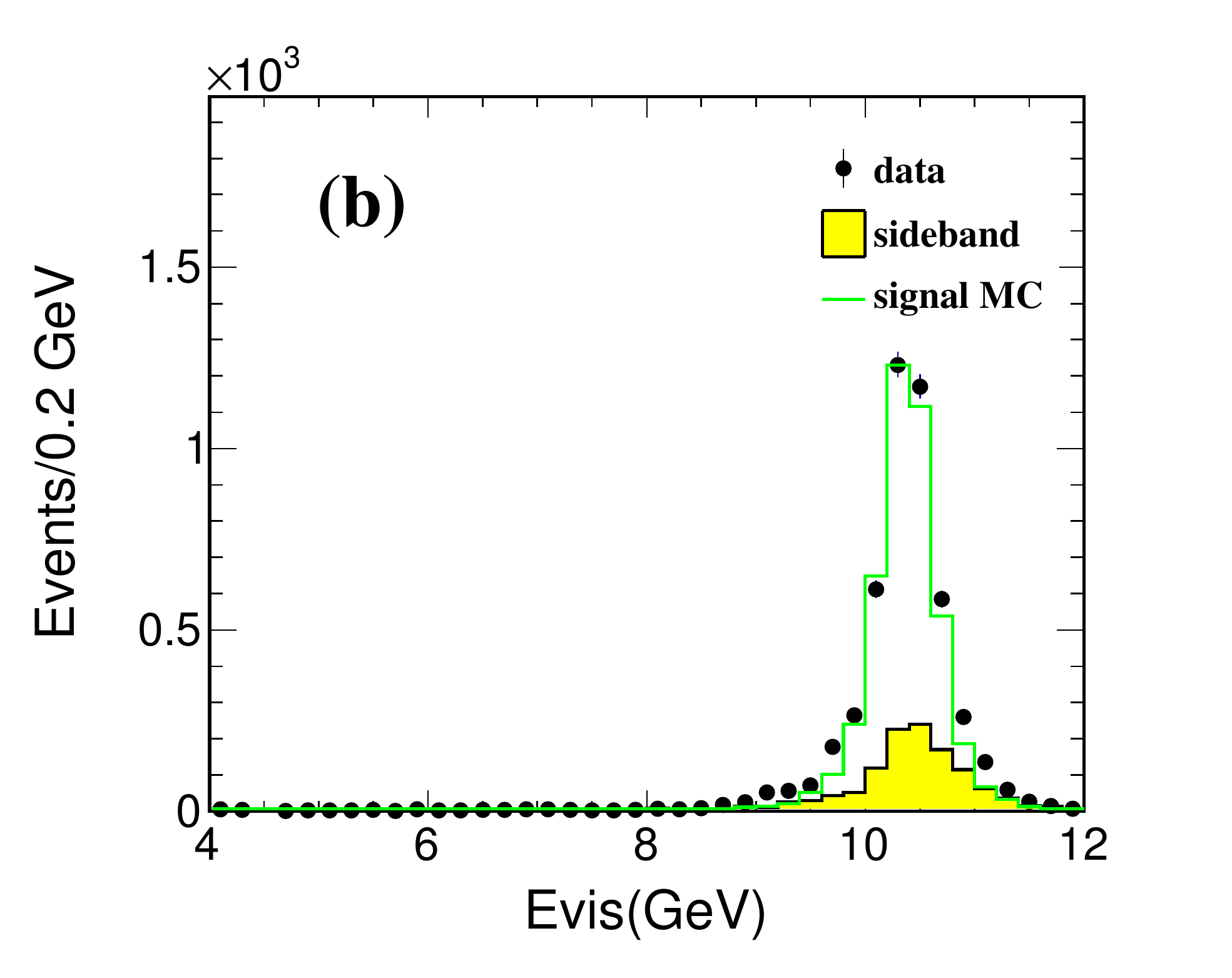}
\includegraphics[width=0.3\textwidth]{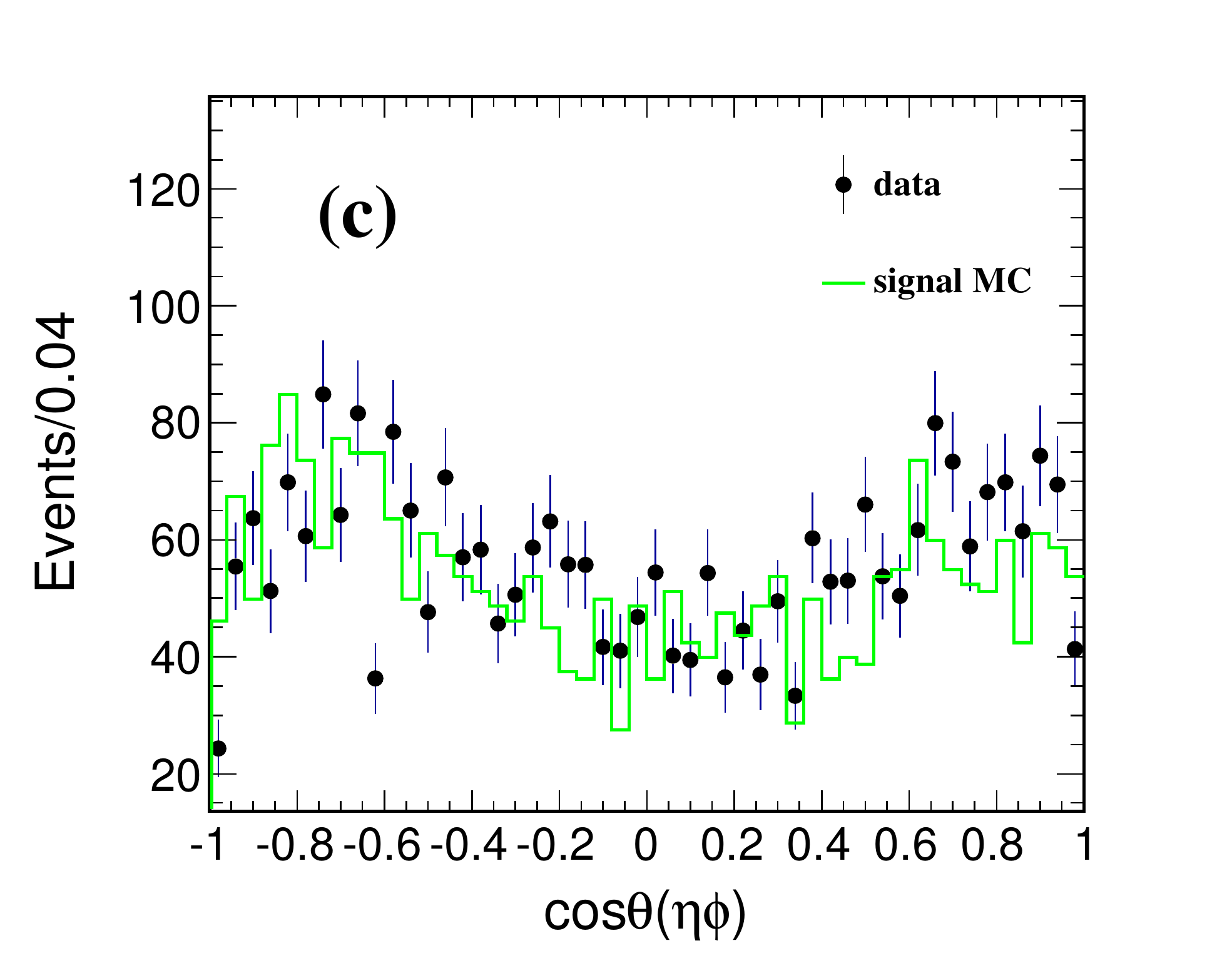}
\caption{The ISR characteristics of the final states. Plot (a) shows the missing mass squared of $\eta\phi$ and $\gisr$, 
(b) shows the visible energy in the detector and (c) shows the angular distribution of $\eta\phi$ in the $\EE$ CM frame. 
The dots with error bars correspond to data while the shaded histograms correspond to backgrounds estimated from the 2D 
sidebands. The unshaded histograms are the signal MC simulations. In plot (c), the backgrounds estimated from 2D 
sidebands have been subtracted from the data.} 
\label{ISR}
\end{figure}

For most of the ISR events, the missing mass squared of the reconstructed $\eta$, $\phi$, and $\gisr$ candidates 
($\MMS(\gisr\eta\phi)$) is close to zero, consistent with either complete reconstruction or a low-energy, second ISR 
photon eluding detection (Fig.~\ref{ISR}(a)). We also require $|\MMS(\gisr\eta\phi)| < 0.1~{\rm GeV^2}/c^4$ with a 
mass-selection efficiency of $(97.7 \pm 0.3)\%$ in the $\ppp$ mode and $(97.1\pm0.3)\%$ in the $\GG$ mode. 
Figures~\ref{ISR}(b) and (c) illustrate the good agreement between data and signal MC simulations for the distributions 
of visible energy of all final state photons and charged particles ($\evis$), as well as the polar angle of the 
$\eta\phi$ system in the $\EE$ CM frame ($\cos\theta(\eta\phi)$), confirming that the signal events are produced via 
ISR.

\section{Invariant mass spectrum of $\eta\phi$ from ISR production}
\label{sec_metaphi}

\begin{figure}[tbp]
\includegraphics[width=0.45\textwidth]{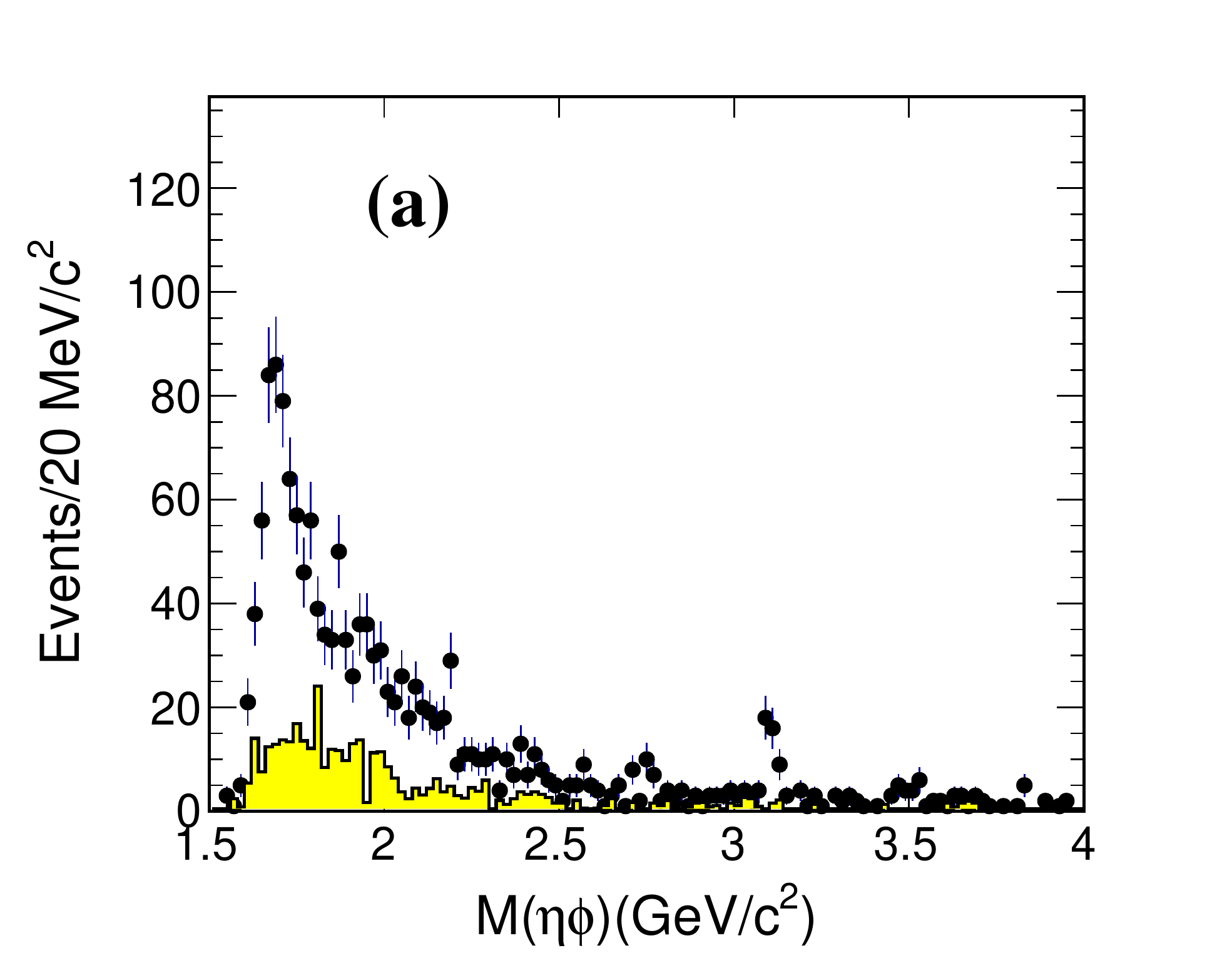}
\includegraphics[width=0.45\textwidth]{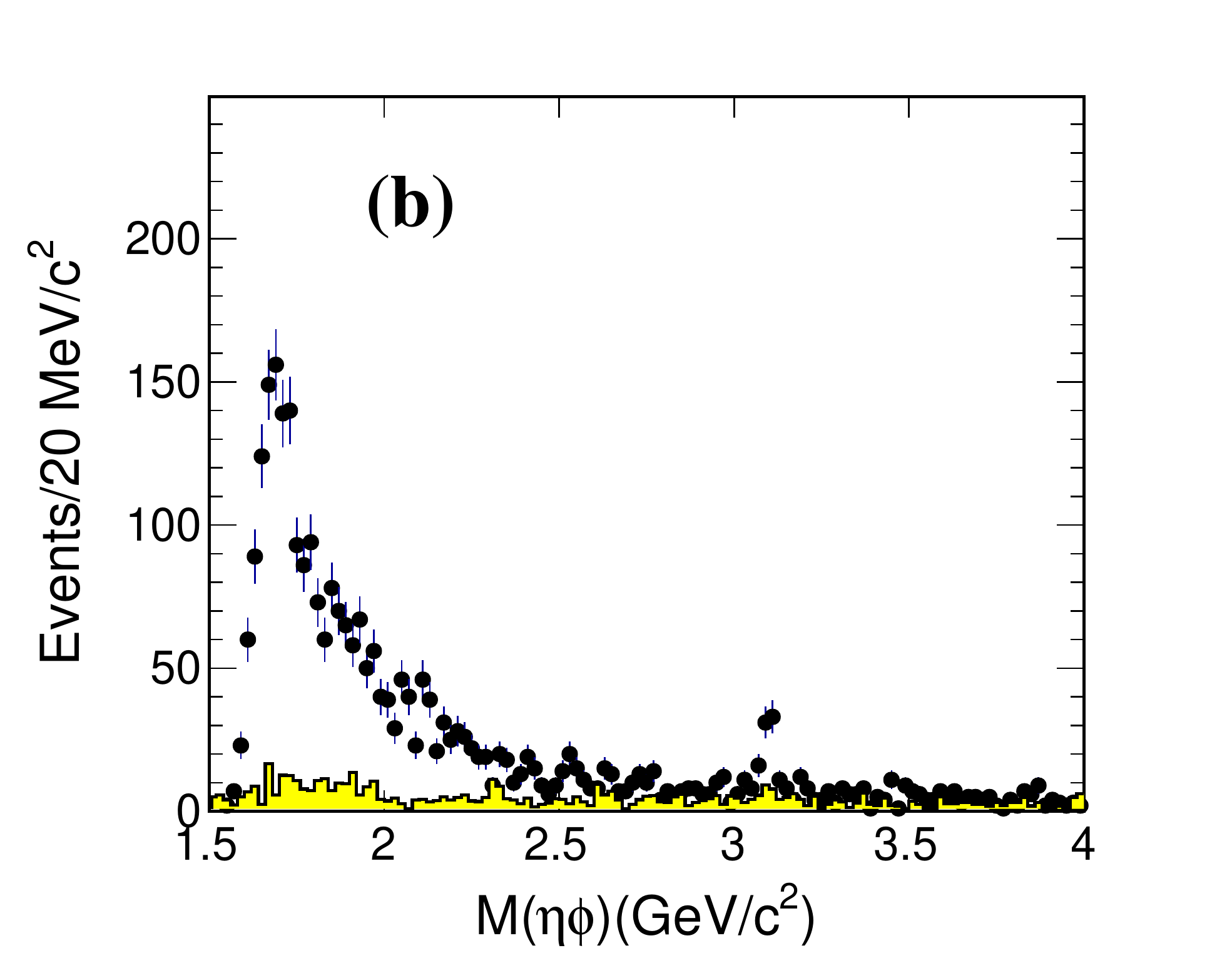}
\caption{Invariant mass $\eta\phi$ distributions in (a) the $\ppp$ mode and (b) the $\GG$ mode from data. The points 
with error bars are from the signal region and the shaded histograms are backgrounds estimated from the 2D sidebands.} 
\label{metaphi}
\end{figure}

After imposing the selection criteria, the distributions of the $\eta\phi$ invariant mass ($M_{\eta\phi}$) from the two 
modes are shown in Fig.~\ref{metaphi}, together with the backgrounds estimated from the scaled 2D sidebands. Using 
$M_{\eta\phi} \equiv M_{\ppp\kk} - M_{\ppp} - M_{\kk} + m_\eta + m_\phi$ for the $\eta\to \ppp$ mode and $M_{\eta\phi} 
\equiv M_{\GG\kk} - M_{\GG} - M_{\kk} + m_\eta + m_\phi$ for the $\eta\to \GG$ mode, the mass resolution of $\eta\phi$ 
is about $6~\mevcs$. The number of obtained $\eta\phi$ signal events is about seven times larger than the previous 
work~\cite{etaphi_babar}, although there is not an obvious $\phi^{''}$ signal.

There are clear $\jpsi$ signals in both the $\ppp$ mode and the $\GG$ mode. Performing an unbinned maximum likelihood 
fit to the combined $M_{\eta\phi}$ spectrum of the two modes, with a Gaussian function for the $\jpsi$ signals and a 
second-order polynomial function for the backgrounds. The $\jpsi$ signal yield is $N_{\rm sig}^{\rm fit} = (99\pm 14)$. 
To estimate the fitting systematic error, polynomial functions of either first- or third-order are also used for the 
background parameterization. The branching fraction for the $\jpsi\to\eta\phi$ decay is calculated using 
\beq\label{eq_2}
\mathcal{B}(\jpsi \to \eta\phi) = \frac{N_{\rm sig}^{\rm fit}}{\sigma^{\rm prod}_{\rm ISR}\times \mathcal{L}\times\eff 
\times \BR(\phi\to\kk)\times \BR(\eta\to\GG/\ppp)},
\eeq
where $\mathcal{L}$, $\eff$, $\BR(\phi\to\kk)$, $\BR(\eta\to\GG/\ppp)$ are the integrated luminosity of the Belle data 
sample, the detection efficiency, the $\phi\to\kk$ branching fraction, and the combined branching fraction for the 
$\eta\to\GG$ and $\ppp$ final states~\cite{PDG}, respectively; $\sigma^{\rm prod}_{\rm ISR}(\jpsi) = 37.5~\pb$ is the 
cross section for $\jpsi$ production via ISR for the Belle experiment~\cite{y2175_belle}. With systematic uncertainties 
as described below in Sec.~\ref{sys_err}, the branching fraction of $\jpsi\to\eta\phi$ is measured to be $(7.1 \pm 1.0 
\pm 0.5) \times 10^{-4}$, which agrees well with the world average value of $(7.4\pm 0.8)\times 10^{-4}$~\cite{PDG}.

We observe a clear $\phit$ signal in the $\eta\phi$ final state. However, the $\phiy$ is not as prominent as in the 
previous BESIII~\cite{etaphi_bes3} analysis. An unbinned maximum likelihood fit is performed to the $M_{\eta\phi}$ mass 
spectra $\in [1.55, 2.85]~\gevcs$ using signal candidate events and 2D sideband events, simultaneously. Similar to the 
parametrization in BaBar's measurement~\cite{etaphi_babar}, the parametrization for the cross section of 
$\EE\to\eta\phi$ at $\sqrt{s}$ takes the form
\beq\label{eq_1}
\sigma_{\eta\phi}(\sqrt{s}) = 
12\pi\mathcal{P}_{\eta\phi}(\sqrt{s})|A_{\eta\phi}^{n.r.}(\sqrt{s})+ A_{\eta\phi}^{\phit}(\sqrt{s}) + 
A_{\eta\phi}^{\phi(2170)}(\sqrt{s})|^2,
\eeq
where $\mathcal{P}_{\eta\phi}$ is the phase space of the final state, $A_{\eta\phi}^{n.r.}(\sqrt{s}) = a_{0}/s^{a_{1}}$ 
describes the non-resonant contribution (mainly due to the tails of resonances below threshold), and 
$A_{\eta\phi}^{\phit}$ ($A_{\eta\phi}^{\phiy}$) is the $\phit$ ($\phi(2170)$) amplitude. The $\phit$ resonance amplitude 
is described by a Breit-Wigner (BW) function
\beq\label{eq_2}
A_{\eta\phi}^{\phit}(\sqrt{s}) = \sqrt{\BR^{\eta\phi}_{\phit}\Gamma^{\EE}_{\phit}}\frac{
\sqrt{\Gamma_{\phit}/\mathcal{P}_{\eta\phi}(M_{\phit}^2)}e^{i\theta_{\phit}}}{M_{\phit}^2 - s - 
i\sqrt{s}\Gamma_{\phit}(\sqrt{s})},
\eeq
where $M_{\phit}$, $\Gamma_{\phit}$ and $\Gamma_{\phit}^{\EE}$ are the mass, the total width and the partial width to 
$\EE$ for the $\phit$, respectively. $\BR_{\phit}^{\eta\phi}$ is the branching fraction for $\phit\to \eta\phi$ and 
$\theta_{\phit}$ is the relative phase. As shown in BaBar's measurement ~\cite{etaphi_babar}, several major decays of 
$\phit$ contribute to $\Gamma_{\phit}$, such as $K K^*(892)$ and $\eta\phi$. Since $\BR^{KK^{*}(892)}_{\phi(1680)} 
\approx 2\times\BR^{\eta\phi}_{\phi(1680)}$, the phase space effect of $KK^*(892)$ can not be ignored in describing 
$\Gamma_{\phit}$. Therefore, we take the form as in Ref~\cite{etaphi_babar}: 
\beqar\label{eq_3}
\Gamma_{\phit}(\sqrt{s}) & =&  
\Gamma_{\phit}[\frac{\mathcal{P}_{KK^{*}(892)}(\sqrt{s})}{\mathcal{P}_{KK^{*}(892)}(M_{\phit})}\BR^{KK^{*}(892)}_{ 
\phit} + \frac{\mathcal{P}_{\eta\phi}(\sqrt{s})}{\mathcal{P}_{\eta\phi}(M_{\phit})}\BR^{\eta\phi}_{ \phit} \nonumber \\
 & & + (1-\BR^{\eta\phi}_{\phit}-\BR^{KK^{*}(892)}_{\phit})].
\eeqar
Here, $\mathcal{P}_{KK^{*}(892)}$ is the phase space of the $\phit\to KK^{*}(892)$ decay. The other decays of $\phit$ 
are neglected, and their phase space dependence correspondingly ignored. Since both the $KK^{*}(892)$ and the 
$\eta\phi$ contain a vector meson ($V$) and a pseudoscalar meson ($P$), the phase takes the form 
\beq\label{eq_4}
\mathcal{P}_{VP}(\sqrt{s})=[\frac{(s+M^2_{V}-M^2_{P})^2-4M^2_{V}s}{s}]^{3/2}.
\eeq
Since there is no measurement of the $KK^{*}(892)$ final state in this work, we take 
$\BR^{\eta\phi}_{\phit}/\BR^{KK^{*}(892)}_{\phit}$ directly from Ref.~\cite{etaphi_babar}. 

The $A_{\eta\phi}^{\phi(2170)}$ is described by
\beq
A_{\eta\phi}^{\phi(2170)}(s) = 
\sqrt{\BR^{\eta\phi}_{\phi(2170)}\Gamma^{\EE}_{\phi(2170)}}\frac{
\sqrt{\Gamma_{\phi(2170)}/\mathcal{P}_{\eta\phi}(M_{\phi(2170)}^2)}e^{i\theta_{\phi(2170)}}}{M_{\phi(2170)}^2 - s - 
i\sqrt{s}\Gamma_{\phi(2170)}}\cdot\frac{B(p)}{B(p')},
\eeq
where $B(p)$ is the $P$-wave Blatt-Weisskopf form factor and $p$ ($p'$) is the breakup momentum corresponding to the 
$\sqrt{s}$ ($M_{\phi(2170)}$).

The efficiencies of the $M_{\eta\phi}$ signal selection are determined from MC samples generated in the range $1.65 < 
M_{\eta\phi} < 2.8~\gevcs$, and are found to be roughly constant (1.35\%) over this mass interval. The effective 
integrated luminosity of ISR is calculated according to the theoretical prescription from~\cite{kuraev}, corresponding 
to $45~\inpb$ per $10~\mev$ near $1.65~\gev$ and increasing to about $80~\inpb$ per $10~\mev$ near $4.0~\gev$. The 2D 
sideband events from $S_1$, $S_2$ and $S_3$ are described by three Landau functions; exponential functions are 
considered to estimate the systematic uncertainty. 

\begin{figure}[htbp]
\subfigure{
\begin{minipage}{0.6\textwidth}
\begin{flushleft}
\psfig{file=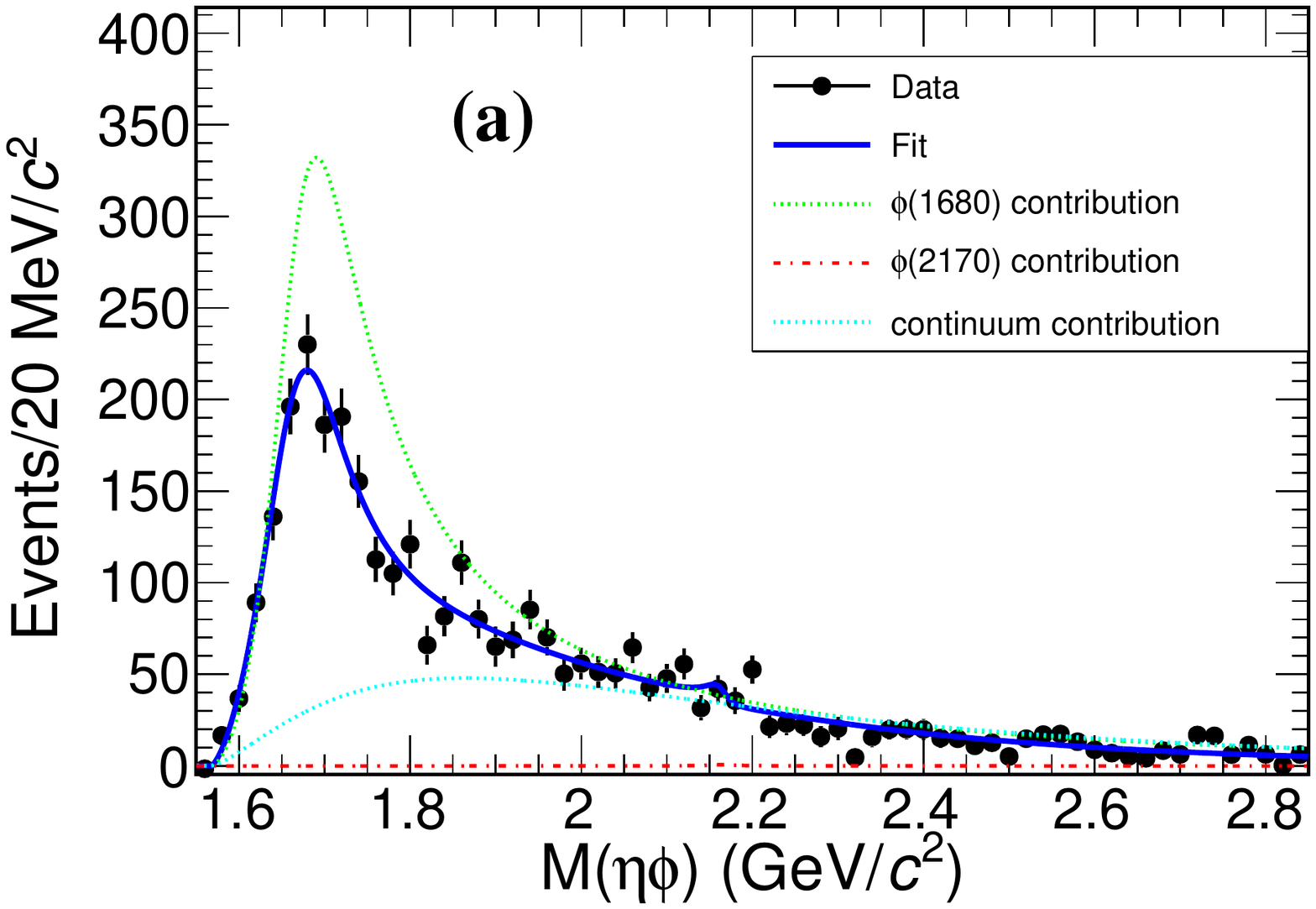, height=3.4cm}
\psfig{file=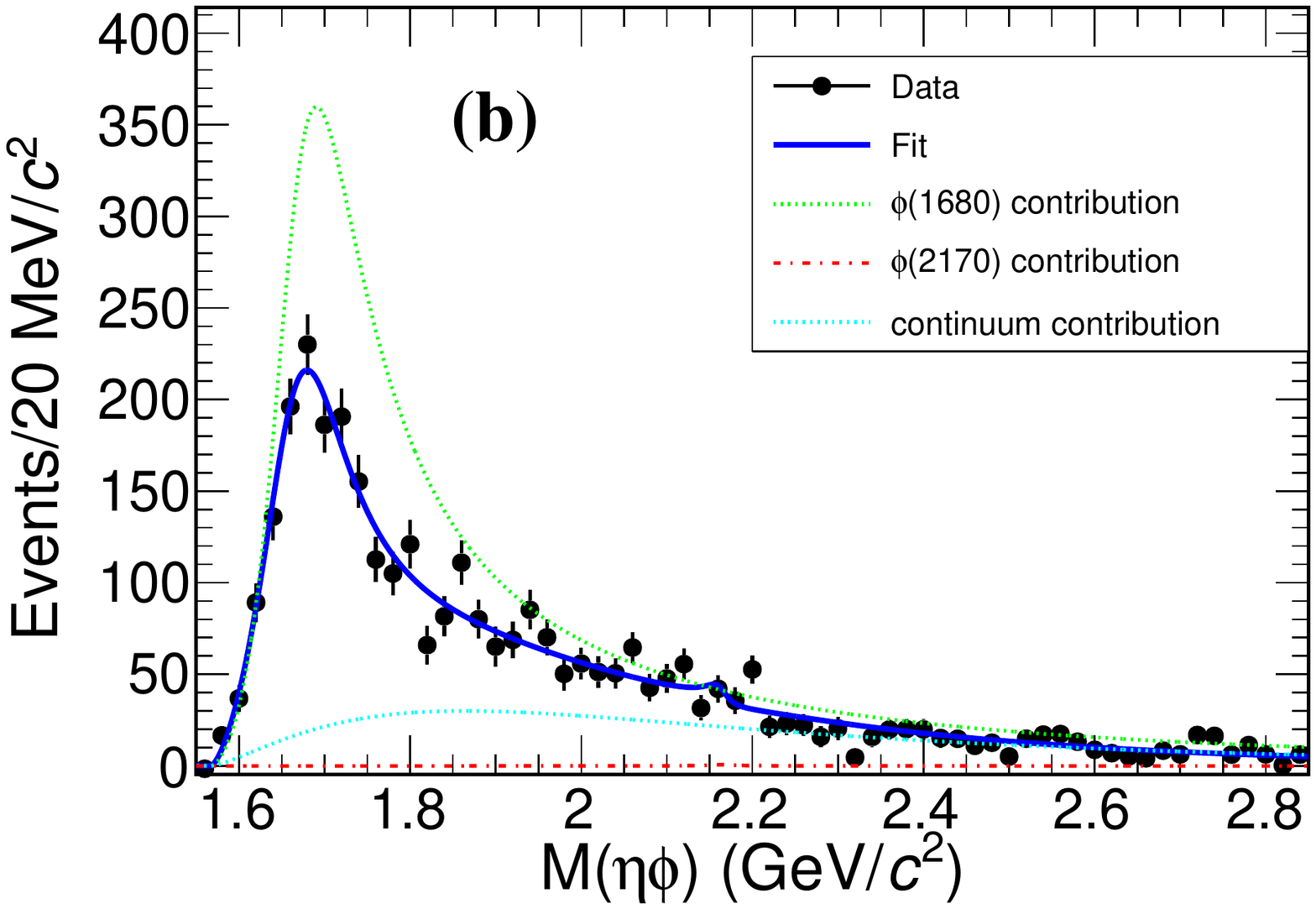, height=3.4cm}\\
\psfig{file=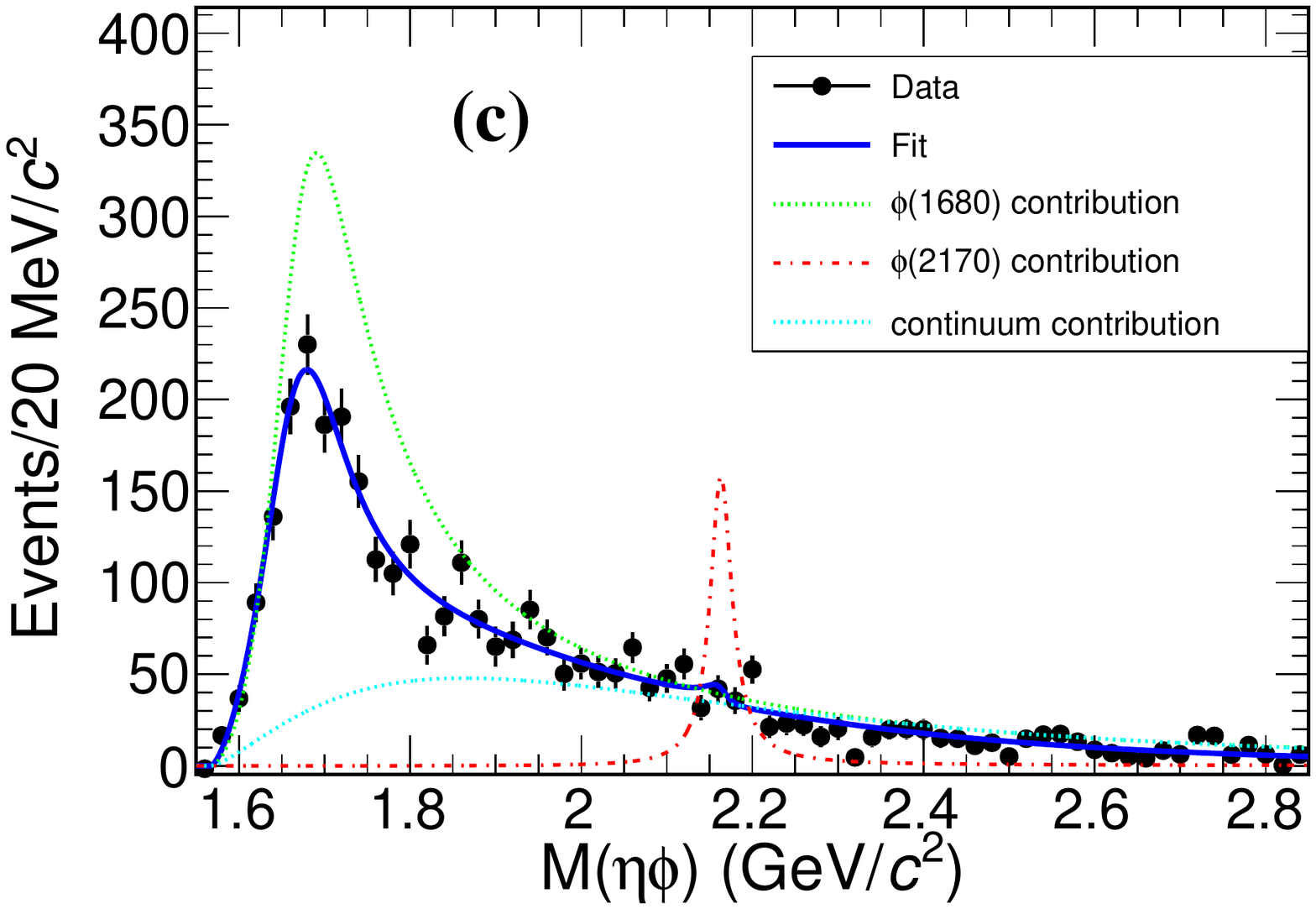, height=3.4cm}
\psfig{file=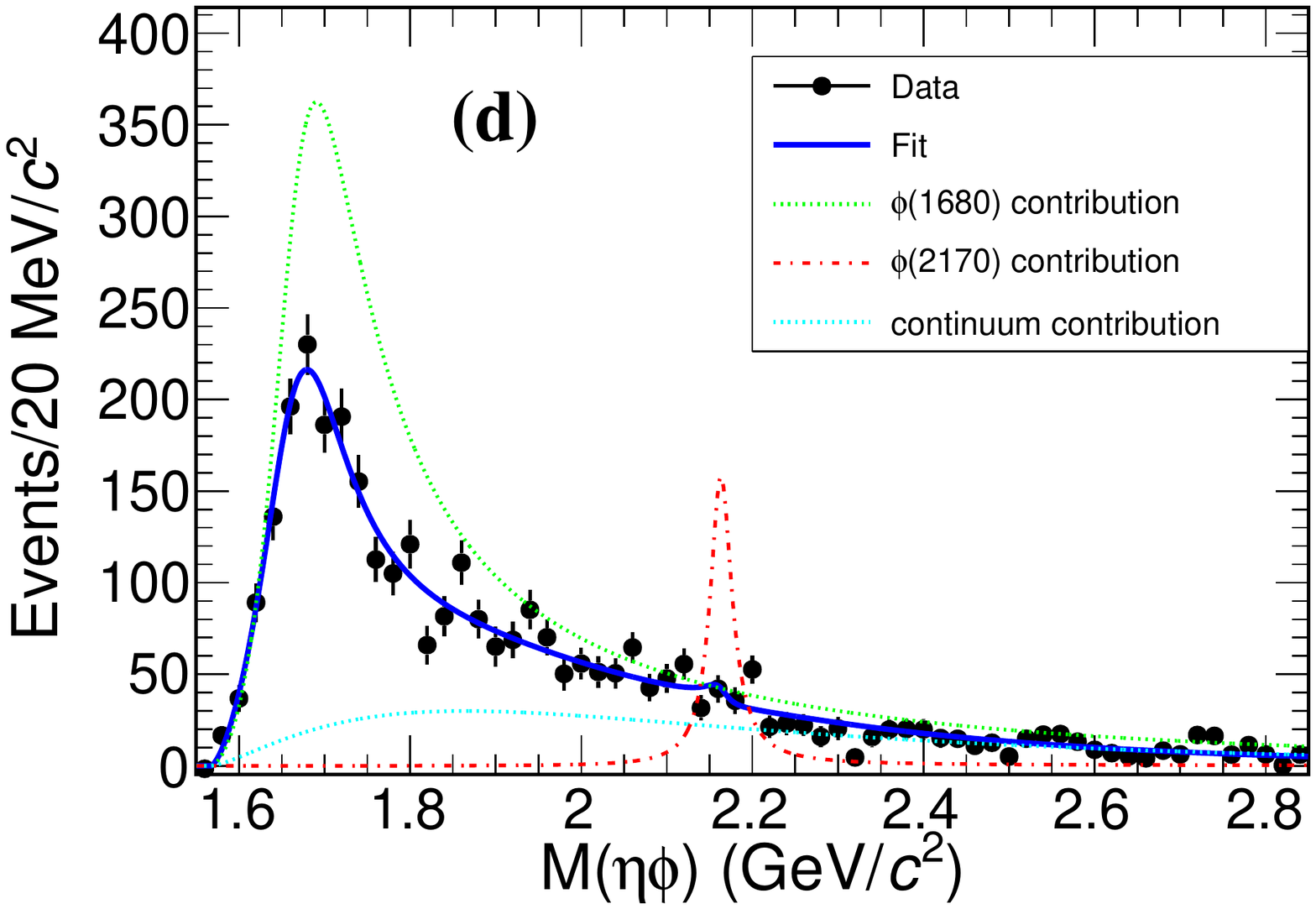, height=3.4cm}
\end{flushleft}
\end{minipage}}
\subfigure{
\begin{minipage}{0.33\textwidth}
\begin{flushleft}
\psfig{file=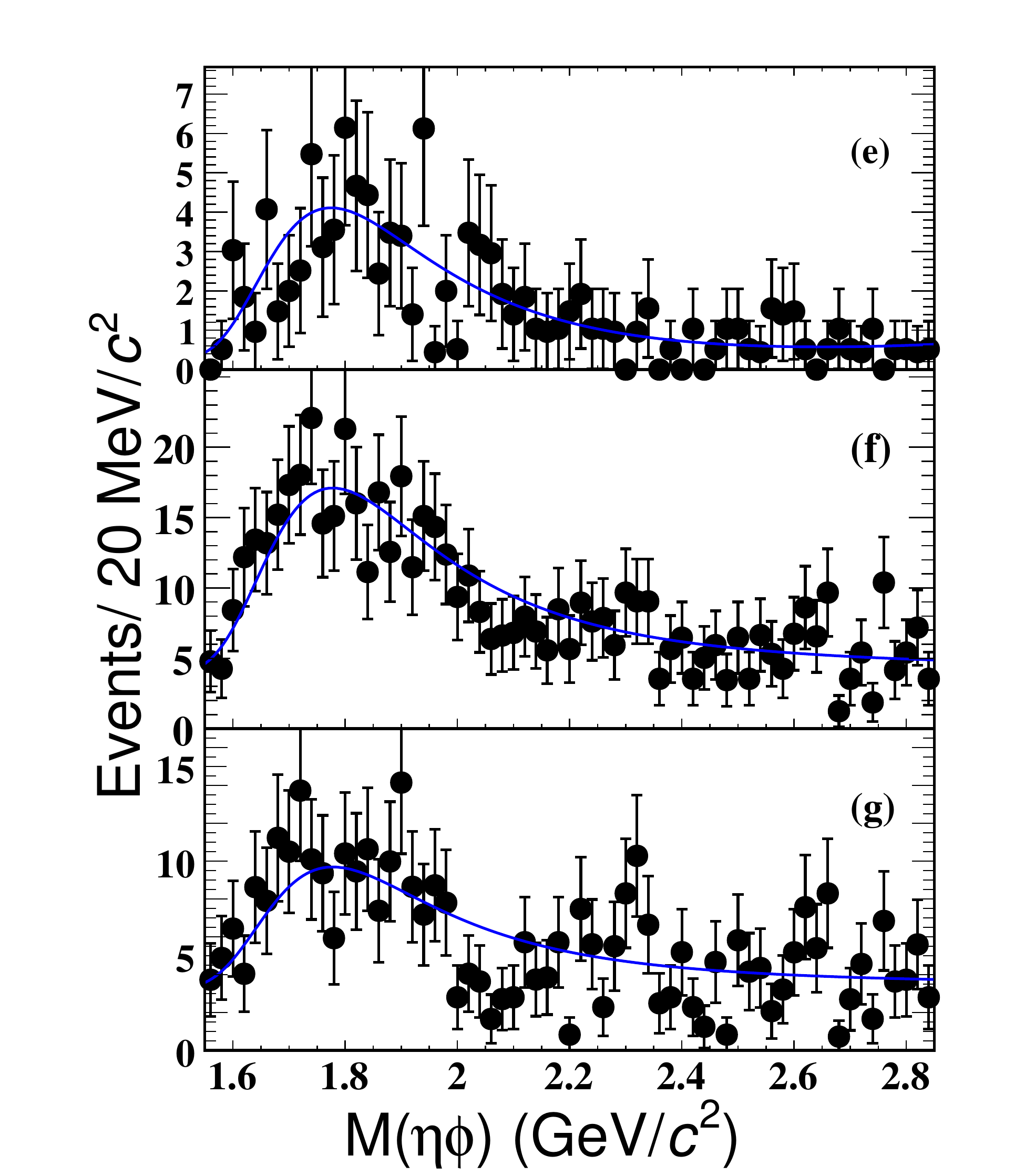, height=7.0cm}
\end{flushleft}
\end{minipage}}
\caption{Invariant mass distribution of $M(\eta\phi)$, and fit results. (a-d) show the four solutions, and (e-g) show 
the backgrounds estimated from 2D sidebands. In (a-d), the backgrounds estimated from 2D sidebands have been 
subtracted. The distribution in (e) shows events from the sideband region $S_3$, (f) from $S_2$ and (g) from $S_1$, 
respectively. The curves show the best fit results, while the interference among continuum, $\phit$ and $\phiy$ are not 
shown.}  
\label{fit}
\end{figure}

Assuming the existence of $\phiy$ in the $\eta\phi$ final state, and fitting using the mass and width of $\phiy$ 
reported by BESIII~\cite{etaphi_bes3}, there are four solutions of equivalent quality, having the same $M_{\phit}$ and 
$\Gamma_{\phit}$. The fit results are shown in Fig.~\ref{fit} and Table~\ref{fit_results}. The reduced chi-squared of 
the fit to the $M_{\eta\phi}$ spectrum is $\chi^2/ndf = 77/56$. The $\phit$ resonant parameters are determined to be 
$M_{\phit} = (1683 \pm 7 \pm 9)~\mevcs$, $\Gamma_{\phit} = (149 \pm 12 \pm 13)~\mev$, and $\BR_{\phit}^{\eta\phi} 
\Gamma^{\EE}_{\phit} = (122 \pm 6 \pm 13) ~\ev$, $(219 \pm 15 \pm 18) ~\ev$, $(163 \pm 11 \pm 13) ~\ev$ or $(203 \pm 12 
\pm 18) ~\ev$ for the four solutions. The branching fraction $\BR_{\phit}^{\eta\phi}$ obtained from the fit is $(18 \pm 
2 \pm 1)\%$, $(19 \pm 4 \pm 2)\%$, $(21 \pm 2 \pm 1)\%$ or $(17 \pm 4 \pm 2)\%$ for the four solutions. The statistical 
significance of $\phiy$ is determined to be $1.7\sigma$ by comparing the value of $\Delta (-2\ln \mathcal{L}) = 
-2\ln(\mathcal{L}_{\rm max}/\mathcal{L}_0)$ and the change in the number of free parameters in the fits, where 
$\mathcal{L}_{\rm max}$ is the likelihood with $\phiy$ and $\mathcal{L}_0$ without $\phiy$. The quantity 
$\BR_{\phiy}^{\eta\phi}\Gamma^{\EE}_{\phiy}$ is determined to be $(0.09 \pm 0.05)~\ev$,  $(0.06 \pm 0.02) ~\ev$, $(16.7 
\pm 1.2) ~\ev$ or $(17.0 \pm 1.2) ~\ev$ in the four solutions. The upper limit for $\phiy$ production at 90\% confidence 
level (C.L.) is determined by integrating the likelihood versus the $\phiy$ yield, with the upper limit degraded by a 
factor of $1/(1-\sigma_{\rm sys})$ to account for systematic uncertainties. (The systematic uncertainties in the fit 
results and $\sigma_{\rm sys}$ are described below in Sec.~\ref{sys_err}.) Finally, the upper limits for 
$\BR_{\phiy}^{\eta\phi}\Gamma^{\EE}_{\phiy}$ are determined to be $0.17~\ev$ (Solutions I and II), or $18.6~\ev$ 
(Solutions III and IV) at $90\%$ confidence level, respectively. Since the $\phiy$ is not significant in our 
measurement, another fit without $\phiy$ in Eq.~\ref{eq_1} is performed, as also indicated in Table~\ref{fit_results}. 
There is no obvious difference in quality between the curves from fits with or without $\phiy$.

\begin{table}[htbp]
\begin{center}
  \caption{Fit results with $\phit$ and $\phiy$ both included, and also excluding $\phiy$.
    The mass and width of $\phiy$ are fixed 
from the prior BESIII measurement~\cite{etaphi_bes3}.}
\begin{tabular}{c|cccc | cc}\hline\hline
Parameters      &  \multicolumn{4}{c|}{with $\phi(2170)$}  & \multicolumn{2}{c}{without $\phiy$}     \\\hline
                & Solution I   & Solution II   & Solution III    & Solution IV  & Solution I & Solution II \\
$\chi^{2}/ndf$  & \multicolumn{4}{c|}{77/56}  & \multicolumn{2}{c}{85/60} \\
$a_{0}$         & $-4.1\pm0.5$ & $5.0\pm0.7$   & $-5.0\pm0.5$    & $-4.8\pm0.2$ & $-3.2\pm0.7$ & $5.0\pm0.1$  \\
$a_{1}$         & $2.7\pm0.1$  & $2.6\pm0.1$   & $2.7\pm0.1$     & $2.6\pm0.1$  & $2.9\pm0.1$  & $2.6\pm0.1$  \\
$\BR_{\eta\phi}^{\phi(1680)}\Gamma_{\EE}^{\phi(1680)} (\ev)$ 
                & $122\pm6$    & $219\pm15$    & $163\pm11$      & $203\pm12$   & $75\pm10$    & $207\pm16$   \\
$M_{\phi(1680)} (\mevcs)$      & \multicolumn{4}{c|}{$1683\pm7$}  & \multicolumn{2}{c}{$1696\pm8$} \\
$\Gamma_{\phi(1680)}(\mev)$    & \multicolumn{4}{c|}{$149\pm12$}  & \multicolumn{2}{c}{$175\pm13$} \\
$\BR_{\eta\phi}^{\phi(1680)}$  
                & $0.18\pm0.02$ & $0.19\pm0.04$ & $0.21\pm0.02$  & $0.17\pm0.04$ & $0.25\pm0.12$ & $0.23\pm0.10$  \\
$\BR^{\phi(2170)}_{\eta\phi}\Gamma_{\EE}^{\phi(2170)} (\ev)$ 
                & $0.09\pm0.05$ & $0.06\pm0.02$ & $16.7\pm1.2$   & $17.0\pm1.2$  & \multicolumn{2}{c}{---} \\
$M_{\phi(2170)} (\mevcs)$    & \multicolumn{4}{c|}{$2163.5(fixed)$}  & \multicolumn{2}{c}{---} \\
$\Gamma_{\phi(2170)} (\mev)$ & \multicolumn{4}{c|}{$31.1(fixed)$}   & \multicolumn{2}{c}{---} \\
$\theta_{\phi(1680)}(^\circ)$ & $-89\pm 2$ & $96\pm 6$ & $-92\pm 1$ & $-86\pm 7$  & $-87\pm 15$ & 
$108 \pm 22$ \\
$\theta_{\phi(2170)}(^\circ)$  & $37\pm 14$  & $-102\pm 11$ & $-167\pm 6$ & $-155 \pm 5$ & 
\multicolumn{2}{c}{---}  \\
\hline\hline
\end{tabular}
\label{fit_results}
\end{center}
\end{table}

\section{Cross section for $\EE\to\eta\phi$}

\begin{figure}[htbp]
\psfig{file=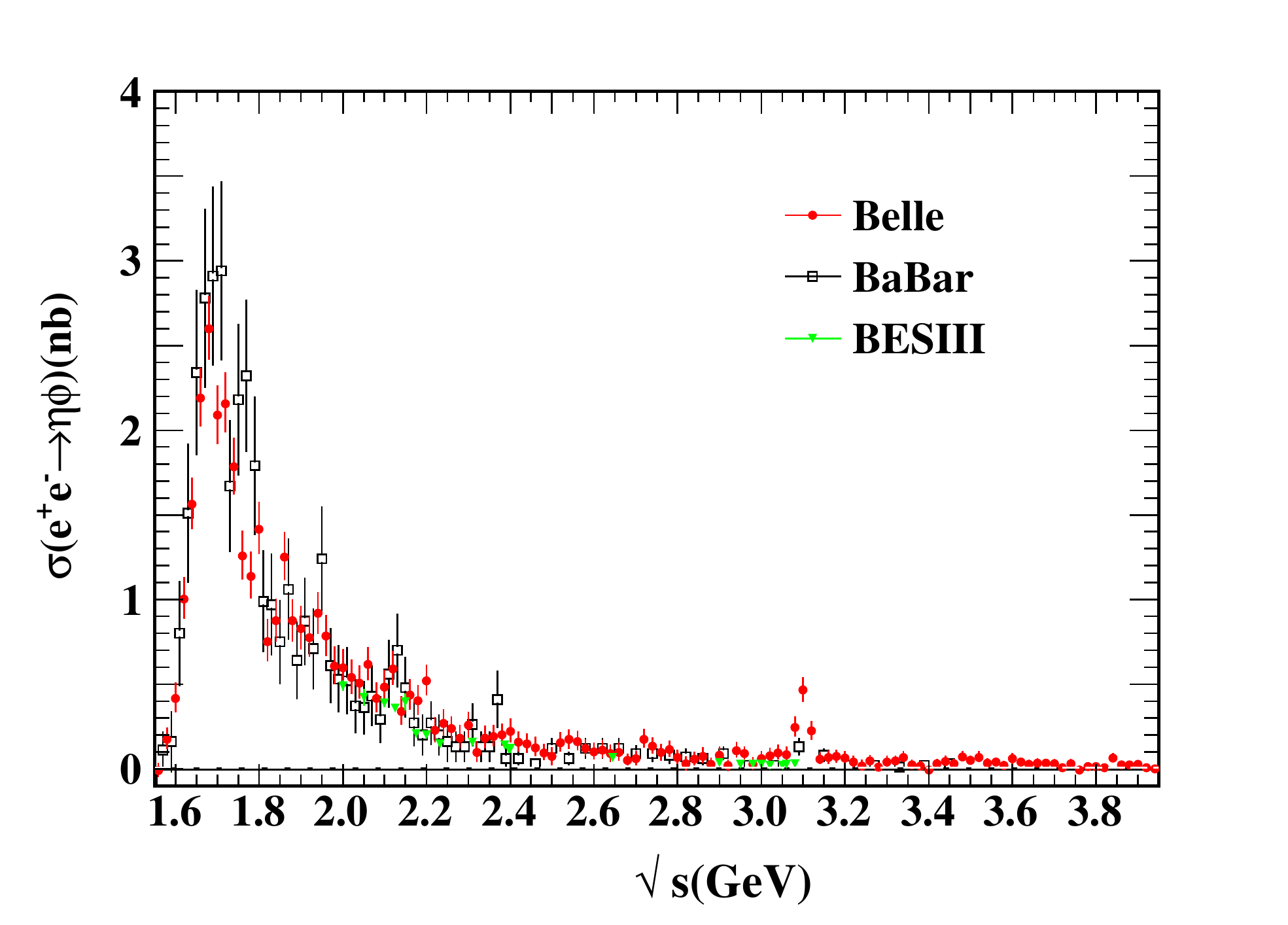,width=0.6\textwidth}
\caption{Cross section for $\EE\to \eta\phi$ from threshold to $3.95~\gev$. The errors are the combination of 
statistical errors and the systematic uncertainties due to the 2D sideband subtraction. A systematic uncertainty of 
6.7\% common to all the data points is not shown.} 
\label{xs_full}
\end{figure}

The $M_{\eta\phi}$ distributions in Fig.~\ref{metaphi} are combined and the cross section of $\EE\to \eta\phi$ for each 
$M_{\eta\phi}$ bin is calculated according to
\beq\label{eq1}
\sigma_i = \frac{n^{\rm obs}_i - n^{\rm bkg}_i}{ \lum_i\times\sum\limits_{j}\eff_{ij}\BR_j},
\eeq
where $i$ is the $i$-th bin of the combined $M_{\eta\phi}$ distribution and $j$ is the $j$-th $\eta$ decay mode; $n^{\rm 
obs}_i$, $n^{\rm bkg}_i$, $\eff_{ij}$, $\lum_i$, and $\BR_j$ are the number of events observed in data, the number of 
background events estimated from the 2D sidebands, the efficiency of signal selection, the effective integrated 
luminosity of ISR production in Belle data, and the branching fractions of $\eta$ and $\phi$ decays~\cite{PDG}, 
respectively. The cross sections for $\EE\to\eta\phi$ measured with Belle data are shown in Fig.~\ref{xs_full}, where 
the error bars include the statistical uncertainties and the systematic uncertainties in the background estimation using 
the 2D sidebands. A 6.7\% common uncertainty (described in Sec.~\ref{sys_err} and Table~\ref{tab7}) is not shown in 
Fig.~\ref{xs_full}. The cross sections for $\EE\to \eta\phi$ are around $2.6~\nb$ and $0.4~\nb$ at the $\phit$ and 
$\phiy$ peaks, respectively. The measured cross section is in good agreement with the results from BaBar's 
measurement~\cite{etaphi_babar}, but with improved precision. 

\section{Systematic uncertainties}
\label{sys_err}

The following systematic uncertainties are characterized for this analysis. The uncertainties due to the particle 
identification are 2.0\% in the $\GG$ mode and 4.0\% in $\ppp$, respectively. The uncertainty due to the tracking 
efficiency is 0.35\% per track and is additive; the uncertainty in the photon reconstruction is 2\% per photon. The 
uncertainties in the $\phi$ mass, $\eta$ mass, and $\MMS(\gisr\eta\phi)$ requirements are measured with the control 
sample $\EE\to \jpsi\to \eta\phi$; $1.3\%$ for the $\eta$ mass window is taken as a conservative uncertainty for the 
combined $\ppp$ and $\GG$ modes. For the $\phi$ mass window, the corresponding value is $0.5\%$. Similarly, $1.3\%$ is 
taken to be a conservative systematic uncertainty estimate, due to the $\MMS(\gisr\eta\phi)$ requirement.

Belle measures luminosity with 1.4\% precision while the uncertainty of the generator {\sc phokhara} is less than 
1\%~\cite{phokhara}. The trigger efficiencies for the events surviving the selection criteria are $(97.0 \pm 0.1)\%$ for 
the $\ppp$ mode and $(95.1\pm 0.1)\%$ for the $\GG$ mode according to the trigger simulation. Conservative uncertainties 
of 1.0\% and 1.5\% are taken to be the systematic uncertainties in the trigger efficiencies for the $\ppp$ mode and 
$\GG$ modes~\cite{y2175_belle, etajpsi_belle}. The uncertainties in the $\phi$ and $\eta$ branching fractions are 
calculated according to the world average values~\cite{PDG}, which contribute a systematic uncertainty of 0.6\%. The 
statistical uncertainty in the MC determination of the efficiency is 0.1\%. 

Assuming all these sources are independent and adding them in quadrature, the total systematic uncertainties in 
measuring $\BR(\jpsi \to \eta\phi)$ are 7.9\% for the $\ppp$ mode and 7.2\% for the $\GG$ mode. There are some common 
uncertainties related to detection efficiency in the two modes, as listed in Table~\ref{tab7}. For other uncertainties 
that have no correlation between two modes, these are first summed in quadrature to obtain $\sigma_{i}$. Then the total 
independent uncertainty ($\sigma_{\rm tot}$) is calculated by $\sqrt{\sum_i (\Delta \eff_i \times \BR_i)^2} / {\sum_i 
(\eff_i \times \BR_i)}$, where $\Delta \eff_i$ equal to $\sigma_i \times \eff_i$, $i$ is $i$th mode of $\eta$ decays ($i 
= \ppp, \GG$). The value of $\sigma_{\rm sys}$ is calculated by $\sqrt{\sum_j (\sigma_j)^2 + (\sigma_{\rm tot})^2}$ 
($\sigma_j$ designates each common uncertainty mentioned above), and the total systematic uncertainty in the cross 
section measurement thereby calculated to be 6.8\%.

By changing the fit range to $[1.6,~2.9]~\gevcs$, the systematic uncertainty due to the fit range is found to be 
negligible. To estimate the model dependence of the non-resonant contribution, we use $A_{\eta\phi}^{n.r.}(s) = 
a_{0}/s$. The uncertainties in backgrounds from the 2D sidebands are estimated by changing $a$ or $b$ by $1\sigma$, and 
changing the functions used to parameterize them, as mentioned in Sec.~\ref{sec_metaphi}. Systematic uncertainties in 
the cross section resulting from different sideband background parameterizations are also shown in Fig.~\ref{xs_full}; 
these translate to uncertainties in the number of $\jpsi$ signal events of 1.8\% in the $\GG$ mode and 1.5\% in the 
$\ppp$ mode. The uncertainty in $\BR^{KK^{*}(892)}_{\phit} / \BR^{\eta\phi}_{\phit}$ is obtained by varying $1\sigma$ 
according to the previous measurement~\cite{etaphi_babar}. 

\begin{table}[htbp]
\begin{center}
\caption{Summary of systematic uncertainties ($\%$) for the measurements of $\BR(\jpsi \to \eta \phi)$ and 
$\sigma(\EE\to\eta\phi)$. Fit uncertainties already described in the text are not included here.}
\begin{small}
\begin{tabular}{c|c|c|c}\hline
Source                &$\GG$ mode & $\ppp$ mode & common  \\\hline
Particle identification & 2.0     & 4.0         & 2.0   \\
Tracking              & 0.7       & 1.4         & 0.7     \\
Photon reconstruction & 6.0       & 6.0         & 6.0     \\
$\phi$, $\eta$ masses and $\MMS(\eta\phi\gisr)$
                      & 1.7       & 1.4         & 1.4     \\
Luminosity            & 1.4       & 1.4         & 1.4     \\
Generator             & 0.5       & 0.5         & 0.5     \\
$\sigma_{\rm ISR}^{\rm prod}(\jpsi)$ 
                      & 1.0       & 1.0         & 1.0     \\
Trigger               & 1.5       & 1.0         & ...     \\
Branching fractions   & 0.6       & 0.6         & 0.6     \\
$\jpsi$ signal fitting & 1.8      & 1.5         & ...     \\
MC statistics         & 0.1       & 0.1         & 0.1     \\\hline 
Sum for $\sigma(\EE\to\eta\phi)$ 
                      & 6.9       & 7.3         & 6.7     \\
Sum for $\BR(\jpsi\to\eta\phi)$
                      & 7.2       & 7.9         & 6.8     \\\hline
\end{tabular}
\end{small}
\label{tab7}
\end{center}
\end{table}
\section{Summary}

In summary, the $\EE \to \eta\phi$ cross sections are measured from threshold to $3.95~\gev$. The branching fraction of 
$\jpsi\to \eta\phi$ is measured to be $(7.1 \pm 1.0 \pm 0.5)\times10^{-4}$, which is in good agreement with the world 
average value~\cite{PDG}. There are four solutions with the same fit quality but different phase angles, obtained from 
fitting the invariant mass distributions of $\eta\phi$ and including both $\phit$ and $\phiy$. The resonant parameters 
of $\phit$ are obtained to be $M_{\phit} = (1683 \pm 7 \pm 9)~\mevcs$, $\Gamma_{\phit} = (149 \pm 12 \pm 13)~\mev$, and 
$\BR_{\phit}^{\eta\phi} \Gamma^{\EE}_{\phit} = (122 \pm 6 \pm 13)~\ev$, $(219 \pm 15 \pm 18)~\ev$, $(163 \pm 11 \pm 
13)~\ev$ or $(203 \pm 12 \pm 18)~\ev$ for the four solutions. The branching fraction for $\phit \to \eta \phi$ is 
determined to be $(18 \pm 2 \pm 1)\%$, $(19 \pm 4\pm 2)\%$, $(21 \pm 2 \pm 1)\%$ or $(17 \pm 4 \pm 2)\%$ for the four 
solutions. We do not find a significant $\phiy$ signal in the Belle data, and instead set an upper limit on its 
production of $\BR_{\phiy}^{\eta\phi} \Gamma^{\EE}_{\phiy} < 0.17~\ev$ or $<18.6~\ev$ at 90\% C.L.; both are consistent 
the BESIII measurement~\cite{etaphi_bes3}. 

\acknowledgments

We thank the KEKB group for the excellent operation of the accelerator; the KEK cryogenics group for the efficient 
operation of the solenoid; 

\end{document}